\documentclass[trackchanges,twocolumn]{aastex701}

\begin{document}

\title{A Universal Physics Defining the Radiation Spectra of Blazars and Gamma-Ray Bursts}

\author{Z. Lucas Uhm}
\altaffiliation{Corresponding author. Email: z.lucas.uhm@gmail.com}
\affiliation{Korea Astronomy and Space Science Institute, Daejeon 34055, Republic of Korea}
\email{z.lucas.uhm@gmail.com}  

\author{Sang-Sung Lee}
\affiliation{Korea Astronomy and Space Science Institute, Daejeon 34055, Republic of Korea}
\affiliation{Astronomy and Space Science Program, University of Science and Technology, Daejeon 34113, Republic of Korea}
\email{fakeemail2@google.com}

\author{Bing Zhang}
\affiliation{Department of Physics, University of Hong Kong, Hong Kong, China}
\affiliation{Nevada Center for Astrophysics and Department of Physics and Astronomy, University of Nevada, Las Vegas, NV 89154, USA}
\email{fakeemail3@google.com}

\author{Judith Racusin}
\affiliation{NASA Goddard Space Flight Center, Greenbelt, MD 20771, USA}
\email{fakeemail4@google.com}

\author{Bindu Rani}
\affiliation{NASA Goddard Space Flight Center, Greenbelt, MD 20771, USA}
\affiliation{Center for Space Science and Technology, University of Maryland, Baltimore County, MD 21250, USA}
\email{fakeemail5@google.com}

\author{Myungshin Im}
\affiliation{SNU Astronomy Research Center, Department of Physics and Astronomy, Seoul National University, Seoul 08826, Republic of Korea}
\email{fakeemail6@google.com}

\author{Aeree Chung}
\affiliation{Department of Astronomy, Yonsei University, Seoul 03722, Republic of Korea}
\email{fakeemail7@google.com}

\author{Hyung Mok Lee}
\affiliation{Center for the Gravitational-Wave Universe, Seoul National University, Seoul 08826, Republic of Korea}
\email{fakeemail8@google.com}



\begin{abstract}
Blazars and gamma-ray bursts (GRBs) are both cosmic beacons of extreme energy release powered by relativistic jets. However, they originate from tremendously different environments. Blazars are the sustained powerhouses driven by supermassive black holes at galactic centers, whereas GRBs are the transient death signals of massive stars or merging compact objects. Here we show that, despite the enormous differences, a universal physics defines the radiation spectra of blazars and GRBs. The blazar spectrum is well described by a ``log-parabola'' function. Employing a simple toy model with a single optically-thin region of a decreasing magnetic field, we produce the log-parabola spectrum very naturally for blazars. We find that the blazar spectrum is shaped by the ``cooling physics'' of relativistic electrons in the fast-cooling regime, which we identify as the universal physics since we previously showed that the fast-cooling physics of electrons with a decreasing magnetic field also explains the mysterious low-energy spectral index of $\gamma$-ray spectrum for a majority of GRBs. This fast-cooling physics of electrons likely nails down the physical origin underlying the universal scaling of the jet energetics between blazars and GRBs, which was observationally suggested more than a decade ago. We highlight that the spectrum shaper in both blazars and GRBs is the cooling physics, not the acceleration mechanism. This finding is conventional-belief-defying and may open up new avenues in a wide range of astrophysics.
\end{abstract}



\section{Introduction}

Blazars and gamma-ray bursts (GRBs) are among the brightest and the most extreme astrophysical phenomena in the Universe. They are both emitted by powerful relativistic jets and appear highly luminous to us due to the relativistic beaming effect since their jets are pointed towards Earth \citep{madau1987,urry1995,kumar2015}. However, they originate from vastly different cosmic environments. Blazars are a type of active galactic nuclei (AGN) that host a supermassive black hole of millions to billions of solar masses at the center of a galaxy \citep{antonucci1993,marscher2008}, whereas GRBs are powered by the formation of a stellar-mass black hole of a few to tens of solar masses \citep{woosley1993,macfadyen1999} or the merger of binary neutron stars \citep{eichler1989,zhangbb2018,lyman2018}. Despite the huge differences in their origin, duration, and the nature of central engines, we show here that a universal physics defines the spectral shape of their synchrotron radiation spectra.

It was widely known that the synchrotron radiation spectrum \citep{rybicki1979} has a spectral index $1/2$ in the fast-cooling regime below the injection energy \citep{sari1998,granot2002}. However, we previously showed that the fast-cooling synchrotron spectrum deviates significantly from this widely-known power-law segment and gets hardened strongly by forming a curved spectral shape when the strength of magnetic fields in the emitting region decreases over time \citep{uhm2014}. This breakthrough finding resulted in a natural explanation for the mysterious low-energy photon spectral index \citep{band1993,preece2000,zhangbb2011} of $\gamma$-ray emission spectrum for a majority of GRBs \citep{uhm2014}.

The blazar spectrum is characterized by a double hump structure \citep{ghisellini1997,fossati1998,chen2014}. Among the two humps, one at lower energies is usually explained by synchrotron emission from relativistic electrons accelerated at shocks \citep{marscher1985,kirk1998,turler1999,bottcher2007}, and its observed spectral shape is well described by a log-parabola function \citep{massaro2004,cheong2024,zuo2025}. Hitherto, this log-parabola shape has been attributed to a single self-absorbed synchrotron spectrum, a possible superposition of multiple slow-cooling synchrotron spectra, or a log-parabola distribution of accelerated electrons \citep{massaro2006,dermer2015}. However, the observed spectral slope at low frequencies below the spectrum peak is far from (i.e., a lot softer than) what is expected in the self-absorption regime \citep{rani2011}, and hence disfavors the self-absorption scenario. Also, in order to form a log-parabola shape out of multiple slow-cooling synchrotron spectra, the superposition needs to be done in a very specific manner, which makes the superposition scenario less favorable. The suggestion that the electrons are accelerated into a log-parabola energy distribution assumes a statistical acceleration mechanism, in which the particle acceleration probability depends upon the particle energy itself \citep{massaro2006}. 

In this study, we produce the log-parabola photon spectrum very naturally for blazars, based on a simple physical modeling that considers a single emitting region in an optically-thin regime. The log-parabola shape in our numerical models is generated without the need of any self-absorption, superposition of multiple spectra, or statistical acceleration mechanism.

\section{Our toy model}

Here we describe our toy model. We consider a blazar located at a redshift $z = 1$, which has a conical jet with a half-opening angle $\theta_{\rm jet} = 1^{\circ}$. The jet is launched from its base, and the axis of the jet makes an angle $\theta_{\rm v} = 5^{\circ}$ with our line of sight. An emitting region of the blazar moves along the jet axis with a bulk Lorentz factor $\Gamma = 10$ (measured in the lab. frame where the jet base is at rest). These values are typical for blazars as inferred from observations \citep[e.g.,][]{weaver2022,cheong2024}. The emitting region has a distance $r$ from the jet base at time $t$ in the lab. frame. The magnetic field strength in the rest frame of the emission region, which we call the co-moving frame, decreases following $B(r) = B_0 (r/r_0)^{-b}$, where $B_0$ is a normalization constant, $b$ is a power-law index, and $r_0 = 10^{19}$~cm. The relativistic electrons are accelerated into a power-law distribution of a slope $p=2.4$ above a minimum injection Lorentz factor $\gamma_m$ and continuously injected into the emission region at an injection rate $R_{\rm inj}$ (see Appendix \ref{section:appendixA} for details). 

We do not consider any other physical quantities, nor do we specify a particular energy dissipation mechanism or particle acceleration mechanism. In this way, we keep the model simple and generic. For instance, the so-called ``shock-in-jet'' model \citep{marscher1985}, which is widely discussed in the field of blazars, can be a subset of our toy model. The power-law electron injection in our model can also be attributed to the magnetic reconnection \citep[e.g.,][]{zhang_hao2023}.

The accelerated electrons in the emission region undergo radiative and adiabatic cooling. For radiative cooling, we include both synchrotron radiation and the inverse Compton (IC) scattering. We split the electron injection into infinitesimally small divisions in time space and also in the electron energy space. We individually track the cooling of each infinitesimal group and compute the instantaneous electron spectrum $f(\gamma_e)$ of the emission region (see Appendix \ref{section:appendixA} for details). Here, $\gamma_e$ is the Lorentz factor of electrons. For each infinitesimal group of electrons, we calculate its synchrotron emission spectrum while taking into account the full synchrotron function $F(x)$ \citep{rybicki1979}. Adding up the contributions from all groups, we compute the observed spectral flux density $F_{\nu}$ emitted by the emission region (see Appendix \ref{section:appendixB} for details). 

\section{Results}


\begin{figure*}[t!]
\plotone{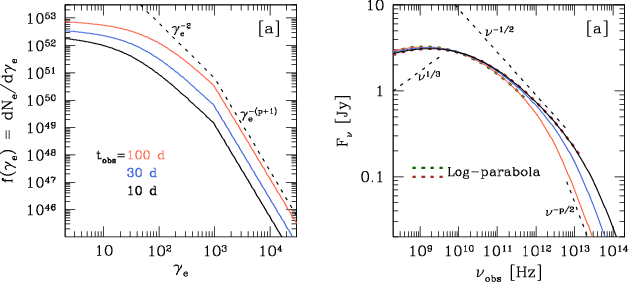}
\caption{
Evolution of the fast-cooling electron-energy spectrum (left panel) and the synchrotron emission spectral flux-density spectrum (right panel) as a function of observer time for the model [a].
\label{fig:f1}}
\end{figure*}

We present six numerical models. In the first model (model [a]), we take the typical values of $B_0 = 200$ mG and $\gamma_m = 10^3$, the index $b=1$ that corresponds to the case of free conical expansion with magnetic flux conservation, and a constant value of $R_{\rm inj} = 10^{47}~{\rm s^{-1}}$ for the electron injection rate. Figure \ref{fig:f1} shows the calculation results for model [a]. The left panel shows the emitting region's electron-energy spectrum $f(\gamma_e)$ at three different observer times $t_{\rm obs}=10$ d (black), 30 d (blue), and 100 d (red): see Appendix \ref{section:appendixC} for details on $t_{\rm obs}$. We first note that the electron spectral slope above the injection Lorentz factor $\gamma_m = 10^3$ maintains the fast-cooling spectral index $p+1$, indicating that the electrons are cooling rapidly in the fast-cooling regime. The electron injection occurs only above $\gamma_m$. Therefore, we emphasize that the electron spectrum below $\gamma_m$ is produced solely by the cooled electrons and thus shaped by the ``cooling physics'' of electrons, by which we mean the physical laws describing the electron energy loss and cooling. This cooling physics is delineated by Equation \ref{eq:gamma_e} in Appendix.

The right panel in Figure \ref{fig:f1} shows the emitting region's synchrotron emission spectrum $F_{\nu}$ at $t_{\rm obs}=10$ d (black), 30 d (blue), and 100 d (red). The photon spectral index approaches the asymptotic slope $p/2$ at high frequencies, which corresponds to the index $p+1$ of the electron spectrum above $\gamma_m$. Apart from this asymptotic behavior, the photon spectrum exhibits a smooth curved shape. This curved shape is emitted by the electron spectrum below $\gamma_m$, which is composed of cooled electrons only. We find that this curved shape is nearly identical to a log-parabola function as shown by the dotted lines. We use a log-parabola function of the form, $\phi (\nu) = \phi_0\, (\nu/\nu_0)^{-\omega \ln (\nu/\nu_0)}$, in which $\phi_0$ gives a normalization at frequency $\nu_0$, and $\omega$ describes how much the spectrum is curved. We plot this log-parabola function over the photon spectrum at 10 d and 100 d with brown and green dotted lines, respectively. As one can see, the photon spectrum in our numerical model is nearly identical to the log-parabola function over up to five orders of magnitude in photon energy. The parameters of the log-parabola function plotted here are summarized in Table~\ref{tab:log-parabola}. Therefore, our calculation here provides a new, natural explanation for the log-parabola shape of blazar spectrum. We highlight that our new explanation holds a clear physical origin: The log-parabola shape is emitted by the electron spectrum portion that is formed solely by cooled electrons and shaped by the cooling physics of electrons.

Not only is it surprising that the log-parabola spectrum is naturally generated based on a simple physical modeling, but it is also very surprising that the generated radio spectrum in model [a] does not evolve much over time, resembling the so-called ``quiescent emission'' of blazars \citep{maraschi1994,acciari2011}. In other words, while the quiescent emission in blazars is usually attributed to the emission from the steady jet itself, now we show that the quiescent emission can also be emitted by an actively emitting region which propagates inside the jet, as shown in model [a].

\begin{figure*}[t!]
\plotone{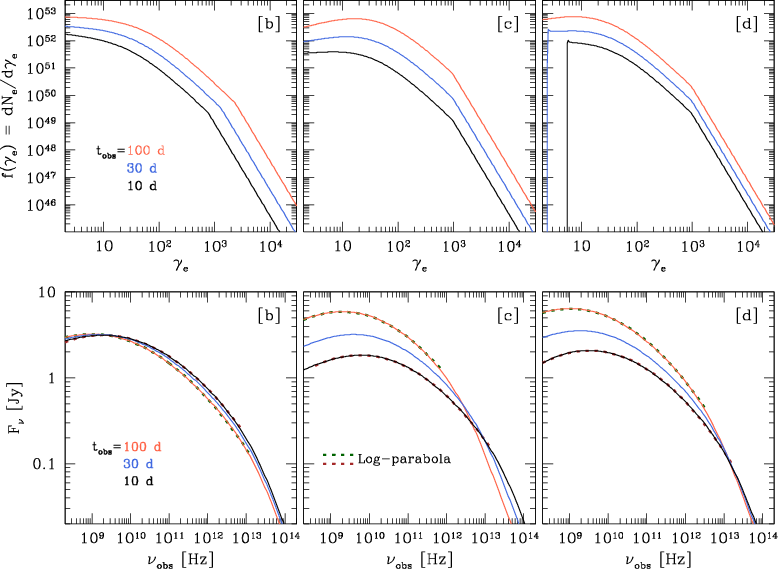}
\caption{
Evolution of the fast-cooling electron-energy spectrum (top row) and the synchrotron emission spectral flux-density spectrum (bottom row) as a function of observer time for the models [b] (left), [c] (middle), and [d] (right column).
\label{fig:f2}}
\end{figure*}

In an attempt to make the radio spectrum to rise over time, we consider three variations on model [a]: (1) Model [b] takes a power-law increase, $\gamma_m (r) = (2 \times 10^3) (r/r_0)^{1/2}$, for the minimum injection Lorentz factor, but all other model setups and parameters are kept the same as in model [a]. (2) Model [c] assumes an increasing profile, $R_{\rm inj}(r) = (2 \times 10^{47}~{\rm s^{-1}}) (r/r_0)^{1/2}$, for the electron injection rate, but besides this, all others remain the same as in model [a]. (3) Model [d] has a shallower decrease, $B(r) = (300~{\rm mG}) (r/r_0)^{-1/2}$, for the magnetic field strength in the emitting region, with everything else kept the same as in model [a].

Figure \ref{fig:f2} shows the calculation results for model [b] (left), [c] (middle), and [d] (right column). The top and bottom panels show their electron spectrum and photon spectrum, respectively, at $t_{\rm obs}=10$ d (black), 30 d (blue), and 100 d (red). In model [b], although the electron spectrum exhibits a rising trend of the injection Lorentz factor over time, the computed spectrum still remains pretty quiescent; rather, the quiescence now extends over an even greater range of photon spectrum than in model [a]. However, in model [c] where the electron injection rate increases over time, the electron spectrum shape shows a substantial change from the case of a constant injection rate, and the photon spectrum rises strongly with its peak frequency decreasing over time. Also, in model [d] where the magnetic field strength falls slower than in other models, we find that the photon spectrum rises significantly with its peak frequency decreasing over time. 

Most importantly, we find that the photon spectrum in all three models [b], [c], and [d] is essentially identical to the log-parabola function over up to five orders of magnitude in photon energy, as one can see from the dotted lines. The parameters of the log-parabola function plotted over the photon spectrum at 10 d and 100 d are shown in Table~\ref{tab:log-parabola}.

We summarize our findings so far: (1) Both the quiescent (model [a] and [b]) and rising (model [c] and [d]) behaviors can be produced out of a single physical setup of our toy model. (2) The quiescent emission is produced when the magnetic field index $b=1$ and the electron injection rate $R_{\rm inj}$ is constant. (3) A single optically-thin region in our model can generate the log-parabola photon spectrum for both quiescent and rising behaviors.

\begin{figure*}[t!]
\plotone{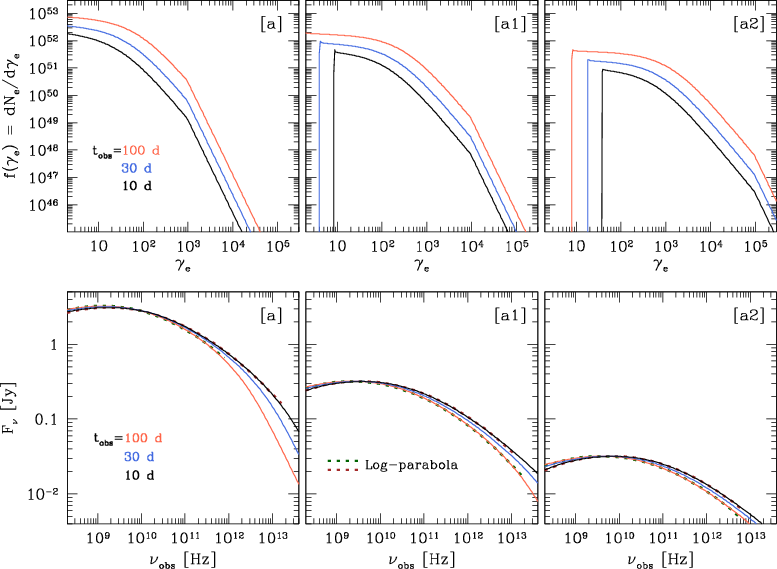}
\caption{
Same as in Figure \ref{fig:f2}, but for the model [a] (left), [a1] (middle), and [a2] (right column).
\label{fig:f3}}
\end{figure*}

\begin{deluxetable*}{lcccccccccccccccccc}
\tabletypesize{\scriptsize}
\tablewidth{0pt} 
\tablecaption{Parameters of log-parabola function plotted in the models [a], [b], [c], [d], [a1], and [a2] for observer times 10 d and 100 d \label{tab:log-parabola}}
\tablehead{
\colhead{} 
& \colhead{} & \multicolumn{2}{c}{Model [a]} 
& \colhead{} & \multicolumn{2}{c}{Model [b]} 
& \colhead{} & \multicolumn{2}{c}{Model [c]} 
& \colhead{} & \multicolumn{2}{c}{Model [d]} 
& \colhead{} & \multicolumn{2}{c}{Model [a1]} 
& \colhead{} & \multicolumn{2}{c}{Model [a2]} \\
\cline{3-4} \cline{6-7} \cline{9-10} \cline{12-13} \cline{15-16} \cline{18-19}
\colhead{Parameters} 
& \colhead{} & \colhead{10 d} & \colhead{100 d} 
& \colhead{} & \colhead{10 d} & \colhead{100 d} 
& \colhead{} & \colhead{10 d} & \colhead{100 d} 
& \colhead{} & \colhead{10 d} & \colhead{100 d} 
& \colhead{} & \colhead{10 d} & \colhead{100 d}
& \colhead{} & \colhead{10 d} & \colhead{100 d}
} 
\startdata 
$\phi_0$ [Jy]    & & 3.1       & 3.28   & & 3.15   & 3.23   & & 1.84   & 5.9     & & 2.07   & 6.37   & & 0.32   & 0.32   & & 0.032 & 0.032 \\
$\nu_0$ [GHz] & & 1.75     & 1.3     & & 1.93   & 0.95   & & 5.66   & 1.9     & & 3.25   & 1.0     & & 3.5     & 2.2     & & 6.9     & 3.8     \\
$\omega$        & & 0.0355 & 0.039 & & 0.037 & 0.036 & & 0.041 & 0.044 & & 0.042 & 0.041 & & 0.034 & 0.036 & & 0.035 & 0.035 \\
\enddata
\tablecomments{We take a log-parabola function of the form, $\phi (\nu) = \phi_0\, (\nu/\nu_0)^{-\omega \ln (\nu/\nu_0)}$, where $\phi_0$ gives a normalization at frequency $\nu_0$, and $\omega$ describes a curvature of the function.}
\end{deluxetable*}

In Figure \ref{fig:f3}, we present two new models [a1] (middle) and [a2] (right column) together with the first model [a] (left column). Model [a1] is identical to model [a] except that $B_0 = 20$ mG and $\gamma_m = 10^4$. Recall that the model [a] has $B_0 = 200$ mG and $\gamma_m = 10^3$. Model [a2] keeps everything the same as in model [a] but has $B_0 = 2$ mG and $\gamma_m = 10^5$. Figure \ref{fig:f3} shows that the log-parabola spectra are also generated in the new models [a1] and [a2] as one can see from the dotted lines. Furthermore, Figure \ref{fig:f3} exhibits a trend that the higher the spectral flux is, the lower the peak frequency is, which is in line with the so-called blazar sequence \citep[e.g.,][]{ghisellini2008}. The parameters of the log-parabola function plotted over the photon spectrum of models [a1] and [a2] are given in Table~\ref{tab:log-parabola}.

\section{Discussion and Conclusion}

The electron spectrum in models [d], [a1], and [a2] has a low-energy cutoff, which marks the current energy of the very first group of electrons injected at $t_{\rm obs} = 0$. The cutoff energy in other models is even lower and unseen. This cutoff energy, which we call $\gamma_c$, is indicative of the ``cooling frequency'' of the emission region, as compared to the ``injection frequency'' that corresponds to the injection energy $\gamma_m$. The fact that $\gamma_c \ll \gamma_m$ in all six models ascertains that the electrons in our models are in the fast-cooling regime: We stress that this is the usual standard fast-cooling regime, not a marginally fast-cooling ($\gamma_c \lesssim \gamma_m$) one. In Appendix \ref{section:appendixD}, we calculate the cooling timescale of the minimum injection Lorentz factor $\gamma_m$, compare it to the dynamical timescale, and show that our models are indeed in the fast-cooling regime.

Although we are in the fast-cooling regime, the electron spectral slope in Figure \ref{fig:f1} below the injection energy $\gamma_m = 10^3$ differs significantly from the widely-known fast-cooling index 2. Also, the photon spectral index of the log-parabola part in Figure \ref{fig:f1} is far from the widely-known fast-cooling index $1/2$. These deviations stem from the fact that the cooling physics in our models involves a decreasing strength of magnetic fields in the emission region over time (or radius) and includes the IC cooling. As shown in our numerical models, this cooling physics actually leads to formation of a smooth curved electron spectrum below the injection energy $\gamma_m$ and to a nearly perfect match to the observed log-parabola spectrum, instead of leading to the widely-known power-laws.

The fast-cooling physics of relativistic electrons in a region of a decreasing magnetic field was also the key factor which enabled us to explain the mysterious low-energy spectral shape of $\gamma$-ray spectrum of GRBs \citep{uhm2014}. Hence, we have identified a common physics which defines the spectral shape of both blazars and GRBs. This fast-cooling physics of electrons is very likely the physics underlying the universal scaling of the jet energetics between blazars and GRBs \citep{nemmen2012}, which was observationally shown more than a decade ago. We suggest the fast-cooling physics of electrons as a universal physics which defines the radiation spectra of relativistic jets like in blazars and GRBs. 

We highlight that the most outstanding finding in this study is: It is not always the acceleration mechanism or the heating physics that leaves its imprints in observations, but it is sometimes simply the cooling physics that we see in observations. This lesson is conventional-belief-defying and may find it applicable in a wide range of high-energy astrophysics, such as micro-quasars, pulsars, X-ray binaries, tidal disruption events, supernova remnants, etc.

In this work, we presented six numerical models to demonstrate the finding that the log-parabola spectra of blazars can naturally arise as a consequence of cooling of relativistic electrons in the fast-cooling regime. However, our exploration of jet parameter space remains limited, and a broader exploration is desired to fully assess the universality of this mechanism across the blazar population. In particular, our calculation of IC cooling parameter considers only the synchrotron self-photons and applies most directly to BL Lacs. In an environment of luminous blazars like FSRQs, the primary seed photons for IC scattering are often external to the jet and thus the external Compton cooling becomes important. A more general treatment implementing such external radiation fields for IC cooling will be presented elsewhere.

\begin{acknowledgments}
We thank the anonymous referee of this paper for helpful and insightful comments. 
This work was supported by the National Research Foundation of Korea (NRF) grant No. 2021M3F7A1084525, funded by the Korean government (MSIT). 
\end{acknowledgments}


\appendix

\section{Cooling physics of relativistic electrons} \label{section:appendixA}

The injection function of electrons takes a power-law distribution
$Q(\gamma_e,t') = Q_0(t') \gamma_e^{-p}$ for $\gamma_e \geq \gamma_m(t')$, where $\gamma_e$ is the Lorentz factor of electrons, $t'$ is the time measured in the co-moving frame, $Q_0$ is a normalization function, $p$ is a power-law index, and $\gamma_m$ is a minimum injection Lorentz factor of electrons. The Lorentz factors $\gamma_e$ and $\gamma_m$ are measured in the co-moving frame. The electrons are continuously accelerated and injected into the emission region at an injection rate 
$R_{\rm inj}(t') = \int_{\gamma_m}^\infty Q(\gamma_e,t') {\rm d} \gamma_e = {\rm d}N_e / {\rm d}t'$, where $N_e$ is the number of accelerated electrons. 

The electrons in the emission region undergo radiative and adiabatic cooling. For radiative cooling, we include both synchrotron radiation and the inverse Compton (IC) scattering. We split the electron injection function $Q(\gamma_e,t')$ into infinitesimally small divisions in time space $t^{\prime}$ and also in the electron energy space $\gamma_e$. The electrons in each infinitesimal group (between $t^{\prime}$ and $t^{\prime}+\delta t^{\prime}$ and between $\gamma_e$ and $\gamma_e+\delta \gamma_e$) share the same injection time $t^{\prime}$ and energy $\gamma_e$. We track the cooling of each group individually and find its current location in the electron energy space. We compute this cooling for all groups at every calculation step and construct the instantaneous electron spectrum, $f(\gamma_e) \equiv {\rm d}N_e/{\rm d}\gamma_e$, of the emission region at any epoch.

The cooling of electrons with the Lorentz factor $\gamma_e$ is described by \citep{uhm2012,uhm2014}
\begin{equation}
\label{eq:gamma_e}
\frac{\rm d}{{\rm d} t^{\prime}} \left(\frac{1}{\gamma_e}\right) = 
\frac{\sigma_T}{6 \pi m_e c}\, {B}^2\, (1+Y) - 
\frac{1}{3} \left(\frac{1}{\gamma_e}\right) \frac{{\rm d} \ln n_e}{{\rm d} t^{\prime}},
\end{equation}
where $\sigma_T$ is Thomson cross section, $m_e$ is the electron mass, and $c$ is the speed of light. The magnetic field strength $B$ and the electron number density $n_e$ are measured in the co-moving frame. The second term on the right-hand side describes the adiabatic cooling of electrons, and for a conically expanding emission region, we take $n_e \propto r^{-2}$, which gives $d \ln n_e = -2 d \ln r$. Here, $r$ is the distance to the emission region from the jet base. The first term on the right shows the radiative cooling, which includes both synchrotron radiation and the inverse Compton (IC) scattering. The IC cooling is calculated employing the parameter $Y = P_{\rm IC}/P_{\rm syn}$, where $P_{\rm IC}$ and $P_{\rm syn}$ are the IC and synchrotron emission power of an electron with the Lorentz factor $\gamma_e$, respectively. Following \cite{sari2001}, we calculate 
\begin{equation}
\label{eq:Y}
Y = -\frac{1}{2}+\sqrt{\frac{1}{4}+\eta \frac{U_e}{U_B}},
\end{equation}
where $U_e$ and $U_B$ are the energy density of the electrons and the magnetic fields, respectively, measured in the co-moving frame of the emission region, and are given by $U_e = E_e/V$ and $U_B = B^2/8 \pi$. The injected electrons have undergone the cooling, and $E_e$ is the sum of all electrons' current energy. In our toy model, the exact shape of the emission region is of no importance: We consider a conical shell with thickness $d$ and approximate its co-moving volume by $V = \pi R^2 d$, since the jet has a small half-opening angle $\theta_{\rm jet} = 1^{\circ}$. We calculate its radius by $R = \pi r \theta_{\rm jet}/180$ and take $d = 3 \times 10^{16}$ cm. The $\eta$ parameter is the fraction of the electrons' energy that was radiated away during the light-crossing time $t^{\prime}_{\rm cross} = {\rm max}(R, d/2)/c$.

\begin{figure*}[t!]
\plotone{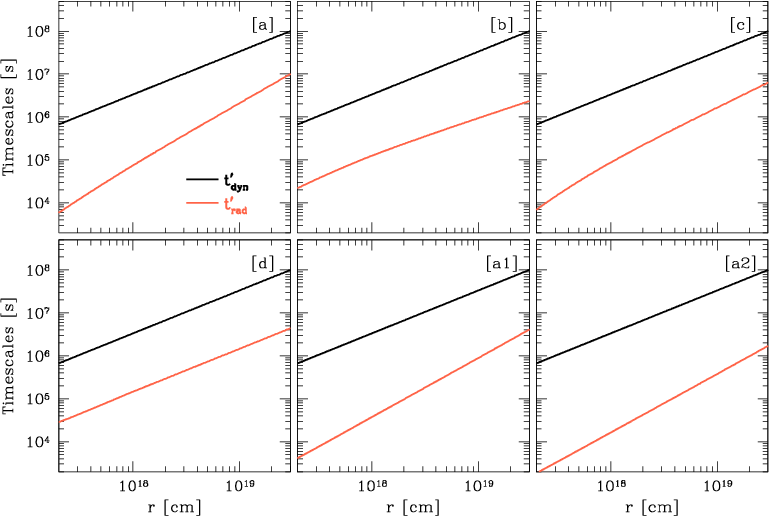}
\caption{
Evolution of the radiative cooling timescale $t_{\rm rad}^{\prime}$ and the dynamical timescale $t_{\rm dyn}^{\prime}$ as a function of the distance $r$ from the jet base for the models [a] (top left), [b] (top middle), [c] (top right), [d] (bottom left), [a1] (bottom middle), and [a2] (bottom right panel).
\label{fig:f4}}
\end{figure*}

\section{Synchrotron emission spectrum} \label{section:appendixB}

For each infinitesimal group of electrons with the current Lorentz factor $\gamma_e$, we calculate its synchrotron emission spectrum while taking into account the full synchrotron function \citep{rybicki1979}, $F(x) = x \int_x^{\infty} K_{5/3}(\xi) {\rm d} \xi$, where $x$ is the ratio of the observed frequency $\nu_{\rm obs}$ to the peak synchrotron frequency of the group, and $K_{5/3}$ is a modified Bessel function of the second kind. Adding up the contributions from all groups of electrons, and taking into account the Doppler boostings between the co-moving frame and the observer frame, we compute the observed spectral flux density $F_{\nu}$ emitted by the emission region. We also include the effects of cosmological expansion on frequencies, spectral fluxes, and observer times. The luminosity distance to the blazar is calculated for a flat $\Lambda$CDM Universe with parameters $\Omega_{\rm m}= 0.315$, $\Omega_{\rm \Lambda}= 0.685$, and $H_0= 67.4$ km/s/Mpc \citep{planck2020}.

\section{Observer time} \label{section:appendixC}

Here we give an expression for observer time $t_{\rm obs}$. The emitting region moves along the jet axis with a bulk Lorentz factor $\Gamma = 10$ and has distance $r$ (from the jet base) at time $t$ in the lab. frame. The axis of the jet makes an angle $\theta_{\rm v} = 5^{\circ}$ with our line of sight. We consider the emission to be turned on at radius $r_{\rm on}$ (and at time $t_{\rm on}$), such that an observer time $t_{\rm obs}$ can be set equal to zero when the first photons from this radius $r_{\rm on}$ are received by an observer on Earth. Then, photons emitted from the emission region at radius $r$ ($>r_{\rm on}$) are received by the observer at
\begin{equation}
\label{eq:tobs}
t_{\rm obs} = \left[ \left(t-\frac{r}{c} \cos \theta_{\rm v} \right) - \left(t_{\rm on}-\frac{r_{\rm on}}{c} \cos \theta_{\rm v} \right) \right] (1+z).
\end{equation}
The time $t$ is calculated by $t=t_{\rm on} + \int_{r_{\rm on}} {\rm d} r/(c\beta)$ where $\beta=(1-1/\Gamma^2)^{1/2}$. Therefore, when $\Gamma$ is constant, Equation \ref{eq:tobs} yields
\begin{equation}
t_{\rm obs} = \left[ \frac{1}{c \beta} (r - r_{\rm on})(1- \beta \cos \theta_{\rm v}) \right] (1+z).
\end{equation}
We take $r_{\rm on} = 10^{17}$ cm for the turn-on radius and $z = 1$ for the redshift of the blazar.

\section{Cooling timescale} \label{section:appendixD}

We calculate here the radiative (i.e., synchrotron + IC) cooling timescale $t_{\rm rad}^{\prime}$ for the minimum injection Lorentz factor $\gamma_m$ of electrons, which is given by 
\begin{equation}
t_{\rm rad}^{\prime} = \frac{6 \pi m_e c}{\sigma_T \gamma_m B^2 (1+Y)}.
\end{equation}
We also compute the dynamical timescale $t_{\rm dyn}^{\prime} = r/(c \Gamma)$ in the co-moving frame and compare it to $t_{\rm rad}^{\prime}$. As shown in Figure \ref{fig:f4}, the radiative cooling timescale $t_{\rm rad}^{\prime}$ is much shorter than the dynamical timescale $t_{\rm dyn}^{\prime}$ in all six models [a], [b], [c], [d], [a1], and [a2]. Therefore, our models are indeed in the fast-cooling regime.





\end{document}